\documentclass[10pt,conference]{IEEEtran}
\IEEEoverridecommandlockouts
\usepackage{cite}
\usepackage{amsmath,amssymb,amsfonts}
\usepackage{algorithmic}
\usepackage{graphicx}
\usepackage{textcomp}
\usepackage{xcolor}
\usepackage{booktabs} 
\usepackage{url}
\usepackage{color}
\usepackage{soul}
\usepackage{booktabs}
\usepackage{graphicx}
\usepackage{multirow}
\usepackage{multicol}
\usepackage{mathrsfs}
\usepackage{amsbsy}
\usepackage{tcolorbox}
\usepackage{balance}
\usepackage{microtype}
\usepackage{array}
\usepackage{longtable}
\usepackage{indentfirst}
\usepackage{pgfplots}
\usepackage{enumitem}
\usepackage{eurosym} 
\usepackage[numbers]{natbib}
\usepackage[framemethod=tikz]{mdframed}
\def\BibTeX{{\rm B\kern-.05em{\sc i\kern-.025em b}\kern-.08em
    T\kern-.1667em\lower.7ex\hbox{E}\kern-.125emX}}

\newcommand{\newtext}[1]{{#1}}

\mdfdefinestyle{mpdframe}{
	frametitlerule=false,
	roundcorner=0pt,
	linewidth=0.5mm
}

\newmdenv[style=mpdframe]{conclusion}

\begin{document}

\title{On Privacy Weaknesses and Vulnerabilities in Software Systems
}

\author{\IEEEauthorblockN{Pattaraporn Sangaroonsilp}
	\IEEEauthorblockA{University of Wollongong \\
		New South Wales, Australia \\
		ps642@uowmail.edu.au}
	\and
	\IEEEauthorblockN{Hoa Khanh Dam}
	\IEEEauthorblockA{University of Wollongong \\
		New South Wales, Australia \\
		hoa@uow.edu.au}
	\and
	\IEEEauthorblockN{Aditya Ghose}
	\IEEEauthorblockA{University of Wollongong \\
		New South Wales, Australia \\
		aditya@uow.edu.au}
}

\maketitle

\begin{abstract}
In this digital era, our privacy is under constant threat as our personal data and traceable online/offline activities are frequently collected, processed and transferred by many software applications. Privacy attacks are often formed by exploiting vulnerabilities found in those software applications. The Common Weakness Enumeration (CWE) and Common Vulnerabilities and Exposures (CVE) systems are currently the main sources that software engineers rely on for understanding and preventing publicly disclosed software vulnerabilities. However, our study on all 922 weaknesses in the CWE and 156,537 vulnerabilities registered in the CVE to date has found a very small coverage of privacy-related vulnerabilities in both systems, only 4.45\% in CWE and 0.1\% in CVE. These also cover only a small number of areas of privacy threats that have been raised in existing privacy software engineering research, privacy regulations and frameworks, and relevant reputable organisations. The actionable insights generated from our study led to the introduction of 11 new common privacy weaknesses to supplement the CWE system, making it become a source for both security and privacy vulnerabilities.
\end{abstract}


\maketitle

\section{Introduction} \label{sec:intro}

Technologies and digitalisation are rapidly emerging in our society. People significantly rely on software applications and computing devices in their daily lives, consequently leaving their traceable digital activities, contributions and communications on those digital devices across the Internet \cite{Statista2021b}. Even when people are not using software, data about their normal life activities may also be collected by software applications through the ubiquity of IoT and GPS devices, surveillance cameras, face recognition apps and so on. Thus, our privacy is under constant threat in this current digital age. In fact, privacy invasions and attacks have been increasing significantly in recent years \cite{Statista, OAIC2021}. For examples, a cyber crime in the U.S. in 2018 exposed 471 million personal records, and a breach of a national ID database in India leaked over 1.1 billion records including biometric information (e.g. iris and fingerprint scans) \cite{Statista2021}. Those incidents and threats raise an urgent need for privacy to be deeply integrated into the development, testing and maintenance of software applications.

Although security and privacy are often discussed together, they are \emph{not} the same \cite{Bambauer2013}. Security often refers to protection against the unauthorised access to software applications and the data they collect and store. On the other hand, privacy relates to protection of the individual rights to their personally identifiable information in terms of how those personal data are collected, used, protected, transferred, altered, disclosed and destroyed \cite{ISO/IEC2011}. For example, security controls are put in place to ensure that only people with credentials have access to a software application in a hospital. However, if anyone with valid credentials can see patient health records using this software, then privacy is not protected. This example demonstrates that security can be achieved without privacy, however security is an essential component for privacy protection.

Cyberattacks, either in the form of security or privacy attacks, are often formed by exploiting \emph{vulnerabilities} or \emph{weaknesses}\footnote{The two terms are often interchangeable. Hereby, we will use vulnerabilities to refer to both of them.} found in software systems. For instance, the infamous WannaCry ransomware attack exploited a vulnerability in Microsoft Windows systems, while the Heartbleed vulnerability in OpenSSL has made millions of websites and online platforms across the world vulnerable to cyberattacks. To prevent similar attacks, efforts have been put into understanding and publicly disclosing vulnerabilities so that developers can identify and fix them in their software applications. These efforts have resulted in the widely-known CWE and CVE systems \cite{CWE, CVE}.

However, there have been very little work (e.g. \cite{Yang2013, Ma2013}) in identifying privacy vulnerabilities. A system which specifically records common privacy vulnerabilities does \emph{not} exist yet. Thus, software developers often rely on the CWE and CVE systems to learn about privacy-related weaknesses and vulnerabilities. However, it is not clear to what extent privacy concerns are covered in those systems, and whether privacy receives adequate attention (which it deserves). To answer these questions, we have collected \emph{all} 922 weaknesses recorded in CWE and 156,537 records registered in CVE to date, filtered out non privacy-related records and further analysed the shortlisted records that are privacy-related. We have found only 41 and 157 privacy vulnerabilities in the CWE and CVE systems respectively. The coverage of privacy-related vulnerabilities in both systems is very limited, only 4.45\% in CWE and 0.1\% in CVE. (\textbf{\underline{Contribution 1}}) 

The next questions we aimed to explore are what privacy threats are covered in those privacy-related vulnerabilities in the CWE/CVE systems and if they are adequately cover the privacy threats raised in both research and practice. To answer these questions, we have conducted an explanatory study on the privacy engineering literature, privacy standards and frameworks (e.g. ISO/IEC 29100), regulations in different jurisdiction, including the European Union General Data Protection Regulation (EU GDPR), California Consumer Privacy Act (CCPA), Health Insurance Portability and Accountability Act (HIPAA), Gramm-Leach Bliley Act (GLBA), the U.S. Privacy Act (USPA) and the Australian Privacy Act (APA), and relevant reputable organisations (e.g. OWASP \cite{OWASP2020} and Norton \cite{Nortona}). This explanatory study informed us to develop a taxonomy of common privacy threats that have been raised in research and practice. The taxonomy is built upon the existing well-known privacy threats taxonomy \cite{Stallings2019}. Multiple raters/coders then examined all 41 and 157 privacy vulnerabilities in the CWE and CVE systems, and mapped them to this taxonomy. The Cohen’s Kappa coefficient, used to measure the inter-rater agreement, was obtained at 0.874 and 0.875 for the CWEs and CVEs respectively, an almost perfect agreement, suggesting the strong reliability of the classification. We found that the existing privacy weaknesses and vulnerabilities reported in the CWE and CVE systems cover only 13 out of 24 common privacy vulnerabilities raised in research and practice. Many important types of privacy weaknesses and vulnerabilities are \emph{not} covered such as improper personal data collection, use and transfer, allowing unauthorised actors to modify personal data, processing personal data at third parties, and improper handling of user privacy preferences and consent. (\textbf{\underline{Contribution 2}})

These actionable insights led to our proposal of 11 new common privacy weaknesses to CWE\footnote{We chose CWE instead of CVE since CVEs specify unique vulnerabilities detected in specific software systems and application, while CWEs are at a more abstract, generic level.}. These new CWE entries cover the areas of privacy threats that have been raised in research and practice but do not exist in CWE yet. To further confirm the relevance and validity of our proposal, we extracted real code examples from software repositories that match with the new CWEs. Our contribution follows the CWE's true spirit of a community-developed list, and will enhance the CWE system to serve as a common language and baseline for identifying, mitigating and preventing not only security but also privacy weaknesses and vulnerabilities. (\textbf{\underline{Contribution 3}})

The remainder of the paper is structured as follows. Section \ref{sec:related-work} discusses related existing work in security and privacy vulnerabilities in software applications. The identification of privacy-related vulnerabilities in CWE and CVE is presented in Section \ref{sec:identifying-privacy-vul}. Section \ref{sec:common-privacy-concerns} discusses the taxonomy of privacy threats and how the privacy-related vulnerabilities in CWE and CVE systems cover those privacy threats. Section \ref{sec:cwe-proposal} presents a new common privacy weakness proposal. The threats to validity of our study are discussed in Section \ref{sec:threats}. We conclude and discuss future work in Section \ref{sec:conclusion}. Finally, we provide the details of a replication package and instructions on how to access it in Section \ref{sec:data-availability}.

\section{Related Work} \label{sec:related-work}

Several systems have been established to standardise the reporting process and structure of common vulnerabilities (e.g. CWE \cite{CWE}, CVE \cite{CVE} and OWASP \cite{OWASP2020}). However, identifying the root causes of the reported vulnerabilities is still a time-consuming and expertise-required process \cite{Gonzalez2019}. Recent work employed information retrieval, data mining, natural language processing, machine learning and deep learning techniques to characterise vulnerabilities reported in CVE and CWE systems \cite{Li2017, Gonzalez2019, Liu2020a}. \newtext{\citeauthor{Li2017} developed a vulnerability mining algorithm to characterise software vulnerabilities \cite{Li2017}. It first created a Vulnerability Knowledge Discovery Database (VKDD) by extracting content from CWE, CVE and National Vulnerability databases. It then employed the semantic model to select terms that are relevant to software vulnerabilities from the VKDD. The association and classification rules were later determined to identify and classify the vulnerabilities. \citeauthor{Liu2020a} built a classification model for detecting vulnerable functions in source code \cite{Liu2020a}. The deep neural network with bidirectional long short-term memory (LSTM) was employed to learn high-level representations of the program's abstract syntax trees. The study also proposed a fuzzy-based oversampling method to mitigate the class imbalance between vulnerable and nonvulnerable code.} However, privacy vulnerabilities were not addressed in those previous work.

Recent work have also studied and made use of the CWE and CVE systems. For example, the work in \cite{Bhandari2021} collected CVE records with their associated CWEs and code commits. The collected information was then analysed to produce insightful metadata such as concerned programming language and code-related metrics. This work can be applied in multiple applications related to software maintenance such as automated vulnerability detection and classification, vulnerability fixing patches analysis and program repair. \citeauthor{Galhardo2020} proposed a formulation to calculate the most dangerous software errors in CWE \cite{Galhardo2020}. They used this formulation to identify the top 20 most significant CWE records in 2019. Again, these prior work only focuses on security vulnerabilities.

\citeauthor{Yang2013} introduced a framework to detect privacy leakage in mobile applications \cite{Yang2013}. This study identified several common privacy vulnerabilities in Android such as unintended sensitive data transmission and local logging. \citeauthor{Ma2013} discussed a privacy vulnerability in mobile sensing networks which collect mobility traces of people and vehicles (e.g. traffic monitoring) \cite{Ma2013}. Although these networks receive anonymous data, it was proven in the study that these data can identify victims. These studies have confirmed the occurrence of privacy vulnerabilities in multiple types of software systems (e.g. web/mobile applications and sensing networks). However, most of the existing studies only focused on security concerns when investigating software vulnerabilities, thus overlooked privacy-related concerns in many contexts.

A number of studies have proposed state-of-the-art taxonomies of privacy threats (e.g. \cite{Solove2006a, Deng2011, Wuyts2014a}). \citeauthor{Solove2006a} proposed four categories of privacy threats which cover harmful activities that can violate privacy of individuals \cite{Solove2006a}. Another well-known standardised taxonomy of privacy threats in software known as LINDDUN was first proposed in \cite{Deng2011}. LINDDUN is a model-based technique used to discover the privacy threats in a system \cite{Wuyts2014a}. It models the system as a data flow diagram (DFD), maps DFD elements to privacy threat categories, and then identifies and documents the privacy threat type and the DFD element type. Although LINDDUN can be applied to any general-purpose software systems, it requires the details of system description to construct the DFD for threat analysis.

\newtext{Two renowned organisations have also worked on privacy framework and privacy threat modelling. National Institute of Standards and Technology (NIST) proposed a privacy framework which focuses mostly on providing high-level guidelines that organisations can follow to govern and control privacy risks \cite{NIST2020}. It does not address the privacy threats specifically in software systems. Thus, we did not include this framework in our study. The MITRE Corporation has been developing a privacy threat modelling method \cite{Bloom2022}. However, the MITRE work took a different approach: they identified privacy attacks from the Federal Trade Commission (FTC) and Federal Communications Commission (FCC) closed cases. By contrast, we identified privacy threats from existing privacy engineering literature, privacy regulations/frameworks, and industry sources. This results in several major differences in the two taxonomies in terms of scope of privacy threats, coverage of privacy threats, and classification of threats and (sub-)categories. In addition, our work goes even beyond providing a taxonomy. We provided a detailed, concrete (CWE-ready) description of the common privacy vulnerabilities including how they occur and their consequences, how to detect them and mitigate them, and demonstrative examples. These provide software engineers with more concrete information of privacy vulnerabilities.}
\section{Identifying privacy-related vulnerabilities in CWE and CVE} \label{sec:identifying-privacy-vul}

\begin{figure}[ht]
	\centering
	\includegraphics[width=1.0\linewidth]{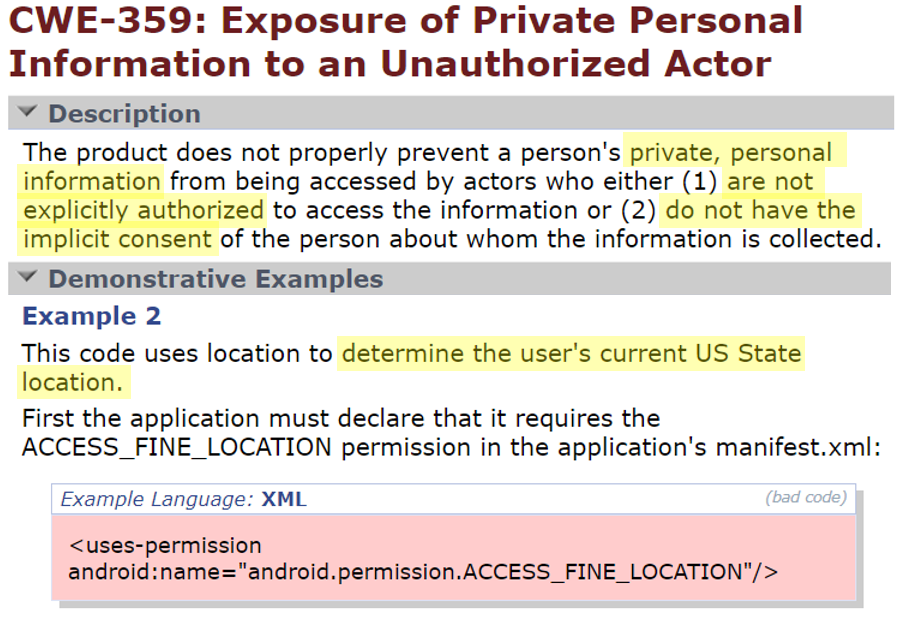}
	\caption{A screenshot highlights privacy-related information in CWE-359.}
	\label{fig:cwe-359}
\end{figure}

Common Weakness Enumeration (CWE) and Common Vulnerabilities and Exposures (CVE) are two well-known systems for publicly known weaknesses and vulnerabilities in software and hardware. 
The CWE system identifies common categories of flaws, bugs and other errors found in software and hardware implementation, code, design or architecture that could be vulnerable to attacks \cite{CWE}. The CWE system has three views: i) by research concepts, ii) by software development and iii) by hardware design. Our study focuses on the research concepts and software development views as they are related to software applications. A CWE record consists of a description, relationships to other CWE records, demonstrative examples, mitigations and other relevant attributes. The interested parties can use this information to identify a weakness in their software systems and applications. For example, CWE-359\footnote{https://cwe.mitre.org/data/definitions/359.html} describes a weakness that exposes private personal information to an unauthorised actor (see Figure \ref{fig:cwe-359}). It also provides demonstrative examples, one of which is a code fragment that exposes a user's location. 

CVE is a catalogue of cybersecurity vulnerabilities that may exist in software products, applications and open libraries (e.g. Skype, Mozilla Firefox and Android). These vulnerabilities are reported by organisations that have partnered with the CVE program. Each CVE record describes the details of a vulnerability and specifies affected version of a software, thus it is more specific comparing to CWE records. For example, CVE-2000-1243\footnote{https://cve.mitre.org/cgi-bin/cvename.cgi?name=CVE-2000-1243} refers to a privacy leak in version 3.04 of Dansie Shopping Cart which sensitive information such as user credentials were sent to an e-mail address controlled by the product developers.

In the absence of a system which specifically records common privacy vulnerabilities, software engineers and other interested parties often rely on the CWE and CVE systems for privacy weaknesses and vulnerabilities (like the one in Figure \ref{fig:cwe-359}). However, both CWE and CVE target at cybersecurity, and although security and privacy are often discussed together, they are not the same. Security vulnerabilities are often exploited by unauthorised access to perform malicious actions in software applications. By contrast, privacy vulnerabilities may lead to violations of the individual rights to their personally identifiable information in terms of how those personal data are collected, used, protected, transferred, altered, disclosed and destroyed. Hence, we have explored to what extent privacy concerns are covered in the CWE and CVE systems, and whether privacy receives adequate attention which it deserves.

\subsection{Approach}

\begin{figure}[ht]
	\centering
	\includegraphics[width=.9\linewidth]{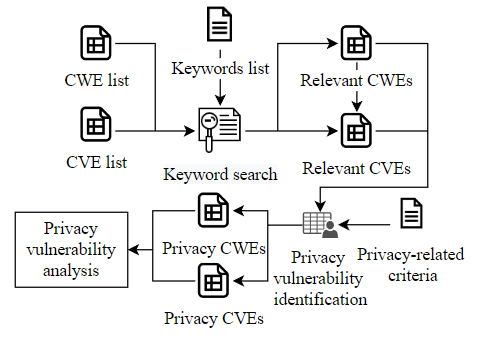}
	\caption{Privacy vulnerabilities identification process.}
	\label{fig:privacy-vul-identification}
\end{figure}

The privacy vulnerability identification process (see Figure \ref{fig:privacy-vul-identification}) consists of the following steps: (i) obtaining the CWE and CVE lists, (ii) determining a list of keywords and performing a keyword search, (iii) identifying privacy-related criteria and annotating the CWE and CVE records, and (iv) performing privacy vulnerability analysis.

To identify the privacy vulnerability in CWE and CVE lists, we first downloaded the whole CWE records in the research concepts and software development views\footnote{The CWE data is available at https://cwe.mitre.org/data/downloads.html} and CVE records\footnote{The CVE data is available at https://cve.mitre.org/data/downloads/index.html} from their websites. We examined all the attributes of 922 weaknesses in both research concepts and software development views in the CWE to date. The CVE list contains 156,537 records to date, all of which were examined in our study. We examined all of the following attributes in each CVE record: name (i.e. CVE-ID), status, description, references, phases, votes and comments.

After obtaining both lists, we then performed keyword searches to filter out the CWE and CVE records that do not have privacy-related keywords. We used a search function in Microsoft Excel to examine the keywords in those records. A set of keywords consists of 37 words categorised into 4 groups as follows:

\begin{itemize}[leftmargin=*]
    \item Group 1: general terms that relate to weaknesses and vulnerabilities in privacy. The terms include privacy, violation, leak and leakage (4 keywords). The term \emph{privacy} generally appears in the CWE and CVE that reported privacy weaknesses and vulnerabilities. On top of that, we also include the terms \emph{violation, leak and leakage} as they are alternatively used to express concerns in the context of privacy (see CWE-359\footnote{https://cwe.mitre.org/data/definitions/359.html} for more details).

    \item Group 2: terms used to refer to personal data or personally identifiable data. They could be sometimes used interchangeably across regulations, standards and industry sources. The keywords in this group include personal information, personal data, sensitive information, sensitive data, private information, private personal information, personally identifiable information, PII, protected health information, PHI, health information and health data (12 keywords). 

    \item Group 3: terms that relate to relevant privacy and data protection regulations/standards/frameworks. In this group, we select a set of specific well-known and widely-adopted data protection regulations and privacy frameworks mentioned in the CWE records, which includes General Data Protection Regulation, GDPR, California Consumer Privacy Act, CCPA, Health Insurance Portability and Accountability Act, HIPAA, Gramm-Leach Bliley Act, GLBA, Safe Harbor Privacy Framework, ISO/IEC 29100 (10 keywords). In addition, we also include the general terms to ensure that we cover unseen regulations and frameworks which are regulation, data protection, privacy act, privacy framework and privacy standard (5 keywords).

    \item Group 4: terms that are frequently seen in the privacy policies and literature when discussing personal data protection and user privacy. These include right(s), consent, opt in/opt-in, opt out/opt-out, preference and breach (6 keywords).
\end{itemize}

We acquired 185 CWE and 1,088 CVE records that contain at least one of the specified keywords. Next, we manually examined each of those records to identify privacy vulnerabilities. A vulnerability is considered as privacy-related if it satisfies one of the following criteria:

\begin{enumerate}[leftmargin=*]
	\item A weakness or vulnerability involves with any processing of personal data (e.g. collection, use, storage, transfer, alteration, erasure and disclosure). 

    \item A weakness or vulnerability which may lead to the violations of the individual rights. To extract the individual rights, we first selected a range of well-established data protection and privacy regulations in different domains such as governments, businesses, healthcare and finance (i.e. EU GDPR, CCPA, HIPAA, GLBA, USPA, APA). These regulations have been widely enacted in country- and regional-level, hence they are well respected by organisations worldwide. In each regulation, we went through each article to look for the individual rights of data subjects/patients/consumers. Once we found the individual right, we added it into our list\footnote{See the file named \emph{``14-rights''} in the data folder in the replication package for the complete list of individual rights and their relevant data protection regulations/privacy acts.}.

\end{enumerate}

We went through the shortlisted 185 CWE and 1,088 CVE records to determine the vulnerabilities that meet the above criteria. In addition, we have found that the National Vulnerability Database (NVD) had done some mapping between CVEs and CWEs. Hence, once we have identified privacy-related CVEs, we used this mapping and applied the above criteria to identify additional privacy-related CWE records.


\subsection{Results}

We identified 41 and 157 privacy vulnerabilities in the CWE and CVE records respectively. The first 28 privacy-related CWE records were found after the keyword search and manual examination steps. The additional 13 privacy-related CWE records were later added after being identified by the privacy-related CVE records. They cover a wide range of privacy concerns in software applications such as missing personal data protection, improper access control, insufficient credentials protection and personal data exposures. These weaknesses and vulnerabilities are not only related to security, but also affect privacy of individuals.


We discuss here a few examples of CWE and CVE records that were classified as privacy vulnerabilities or weaknesses and refer the reader to \cite{rep-pkg-privul} for a full list of them. CWE-359\footnote{https://cwe.mitre.org/data/definitions/359.html} refers to the exposure of private personal information to an unauthorised actor. Private personal information here includes social security numbers, geographical location, financial data and health records. In addition, this CWE also mentions relevant data protection regulations and privacy acts such as GDPR and CCPA.

Another example is CVE-2020-13702\footnote{https://cve.mitre.org/cgi-bin/cvename.cgi?name=CVE-2020-13702} which refers to the rolling proximity identifier used in the Apple/Google exposure notification API beta through 2020-05-29. This vulnerability enables attackers to evade Bluetooth Smart Privacy due to a secondary temporary UID through tracking individual device movement using a Bluetooth LE discovery mechanism. This is a privacy vulnerability since it concerns user's location and relates to the processing of user's location. There are many cases where the reported vulnerabilities are \emph{not} specifically privacy-related. For example, CWE-78\footnote{https://cwe.mitre.org/data/definitions/78.html} enables the attacker to execute arbitrary commands on the operating systems, leading to unauthorised access to operating systems. However, this is not specifically a privacy vulnerability since it does not involve personal data, personal data processing or individual rights.

\begin{conclusion}
	\textbf{The coverage of privacy-related vulnerabilities in both CWE and CVE records is quite limited: 4.45\% in the CWE system and 0.1\% in the CVE system.}
\end{conclusion}
\section{Common privacy threats in software applications} \label{sec:common-privacy-concerns}
This section investigates how those privacy-related vulnerabilities identified in CWE and CVE (see Section \ref{sec:identifying-privacy-vul}) address common privacy threats in software applications. We first discuss a taxonomy of common privacy threats that we have developed based on an explanatory study of the literature. We then report the threats that have not been covered by the existing privacy-related vulnerabilities in CWE/CVE.

\subsection{Explanatory study}

To identify common privacy threats in practice, we performed an explanatory study on the literature of the following three groups: existing privacy software engineering (SE) research, well-established data protection regulations and privacy frameworks, and additional reputable resources. The details of this process are described below.

\subsubsection{Privacy engineering research}
Privacy engineering has attracted an emerging area of research in SE\cite{Gurses2016}. Our study in this area followed a systematic literature review process proposed by \cite{Kitchenham2007a} to retrieve relevant papers and conduct a literature survey. This process consists of three phases: planning, papers selection and extracting \& reporting.
In the planning phase, we defined our study goal which is to identify privacy threats addressed in the privacy engineering research. Our research question is ``What are privacy threats caused by software developers and/or relevant parties that were addressed in privacy engineering research?''. To achieve the study goal, we developed a review protocol to determine the scope of our study. The protocol consists of four tasks: (i) determining a search keyword and time scope, (ii) selecting SE publication venues, (iii) determining exclusion and inclusion criteria, and (iv) determining a set of questions to identify the privacy threats in the papers. In the papers selection phase, we conducted a search process and applied inclusion and exclusion to the retrieved papers. Finally, we analysed and identified privacy threats in the selected papers and reported the results.


\paragraph{\textbf{Search keyword and time scope}}

\begin{figure}[h]
	\centering
	\includegraphics[width=1.0\linewidth]{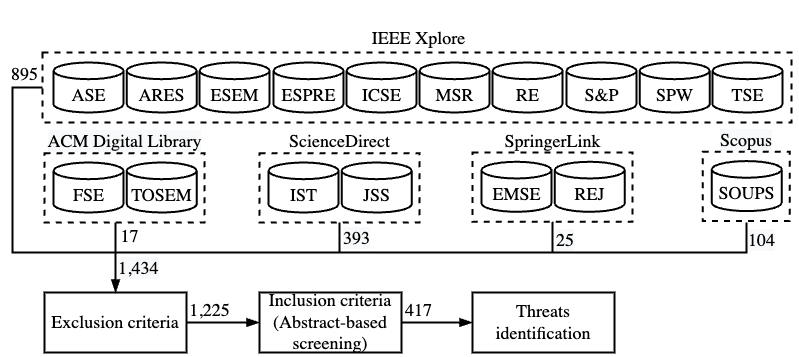}
	\caption{Number of papers included in each step.}
	\label{fig:process-paper-selection}
\end{figure}

We used a search keyword \emph{``privacy''} to search in title, abstract and keywords fields of papers in the selected publication venues. Only one search keyword was used as we already performed a search in the specific SE publication venues, thus we aimed to get all the papers that address privacy. In addition, the papers from these venues are peer-reviewed by experienced researchers in privacy and SE area, hence their significant contributions and quality are well received by the research community. We also determined to search for papers that were published in the past 20 years (2001 - 2020). This task was automatically done by the search functions in the academic databases (i.e. IEEE Xplore, ACM Digital Library, ScienceDirect, SpringerLink and Scopus). We have found 1,434 papers related to privacy during the period of 2001 to 2020 (see Figure \ref{fig:process-paper-selection}).

\paragraph{\textbf{Software engineering publication venues}}

As privacy has also been widely addressed in other fields of study (e.g. law and social sciences), we scope down the search to get a reasonable number of papers in SE by selecting seventeen highly recognised SE publication venues consisting of 7 conferences, 6 journals, 2 symposiums and 2 workshops. These venues focus on various disciplines in SE, which make them proper candidates for representing privacy in multiple areas. Many of these venues were included in the existing papers that conducted systematic literature review in SE area (e.g. \cite{Ebrahimi2019, Perera2020a, Bertolino2018}). The full list of venues, along with the number of papers found in each, is included in the replication package. 

\paragraph{\textbf{Exclusion and inclusion criteria}}

We manually performed an inclusion and exclusion task to ensure that the papers retrieved from the automated search process satisfy our study scope. We initially determined a set of inclusion and exclusion criteria to filter out the papers that are irrelevant to our goal. We exclude the papers that either contain insufficient, incomplete or irrelevant information (e.g. records that are journal/conference/workshop introduction, information from program chairs and summaries of keynotes), are duplicate \emph{OR} are secondary or tertiary studies (e.g. existing systematic literature review papers) and posters. For the inclusion criteria, we include the papers which their contribution is related to privacy \emph{AND} software development.



%

After applying the EC, we excluded 209 out of 1,434 papers. 1,225 papers were passed to the next step. We then applied the IC to the abstracts of the papers. If the papers satisfy both IC, they are included in our study; otherwise they are excluded. Finally, 417 papers were included in our study\footnote{The bibliographic data of those papers are available in \cite{rep-pkg-privul}.}.

\paragraph{\textbf{Privacy threats identification}}

To identify privacy threats, we examine the research studies reported in those papers, and shortlist the ones that focus specifically on privacy vulnerabilities and attacks from this list. The papers were analysed by asking the following questions:

\begin{itemize}[leftmargin=*]
	\item What is a cause of privacy-related problem in software that has been raised in the paper? This cause must involve personal data and is harmful to data subjects. 
	\item Is the identified cause caused by software developers, data controllers/processors, organisations or external parties? This question helps us focus on the privacy threats that are not caused by users. There are papers that investigate user perceptions towards different privacy perspectives (e.g. user perceptions of online behavioural advertising/smart home devices and user confidence in using smartphones). We do not include the privacy threats that are caused by users in this study as it is out of our scope.
\end{itemize}

We manually examined the selected papers, and added privacy threats identified in the papers into a threat list. During the investigation, if we find a threat that is already included in our list, we annotate the paper with the existing threat and proceed to the next paper, otherwise we add a new threat into the list. Finally, we have found 179 papers that address privacy vulnerabilities and attacks in software development\footnote{See the file named \emph{``15-papers-by-threat-categories''} in the data folder in the replication package.}. From these studies, we have identified a range of privacy threats in various applications (e.g. mobile/web applications, mobile sensors, smart home devices, network communications and privacy policy compliance). These will be discussed in details in Section \ref{subsec:areas-of-concern}. The rest of the papers either address the threats caused by users or do not address privacy vulnerabilities or attacks in software systems.



\subsubsection{Privacy regulations and frameworks}

We have studied 7 well-established data protection and privacy regulations and frameworks (i.e. GDPR \cite{OfficeJournaloftheEuropeanUnion;2016}, CCPA \cite{CCPA}, HIPAA \cite{HIPAA}, GLBA \cite{GLBA}, USPA \cite{US1974}, APA \cite{APA} and ISO/IEC 29100 \cite{ISO/IEC2011}). We have found that these regulations and frameworks focus mainly on the privacy threats related to the rights of individuals to control and be informed of their personal data processing. 

\subsubsection{Additional sources}
We have also included a range of relevant reputable organisations (e.g. OWASP and Norton) on this topic. These sources  (e.g. \cite{OWASP2020, Norton}) cover the recent trends of privacy risks and attacks. For example, OWASP identified 20 privacy risks in web applications \cite{OWASPsurvey} such as personal data leak, personal data stolen through common cyberattacks (e.g. cross-site scripting and broken session management). Norton also identified a number of social engineering cyberattacks that are privacy-related such as phishing and keystroke logging attacks \cite{Nortona}.

In the next section, we will discuss in details the common privacy threats that we have found in our explanatory study of the literature.

\subsection{A taxonomy of common privacy threats} \label{subsec:areas-of-concern}

\begin{figure*}
	\centering
	\includegraphics[width=1.0\linewidth]{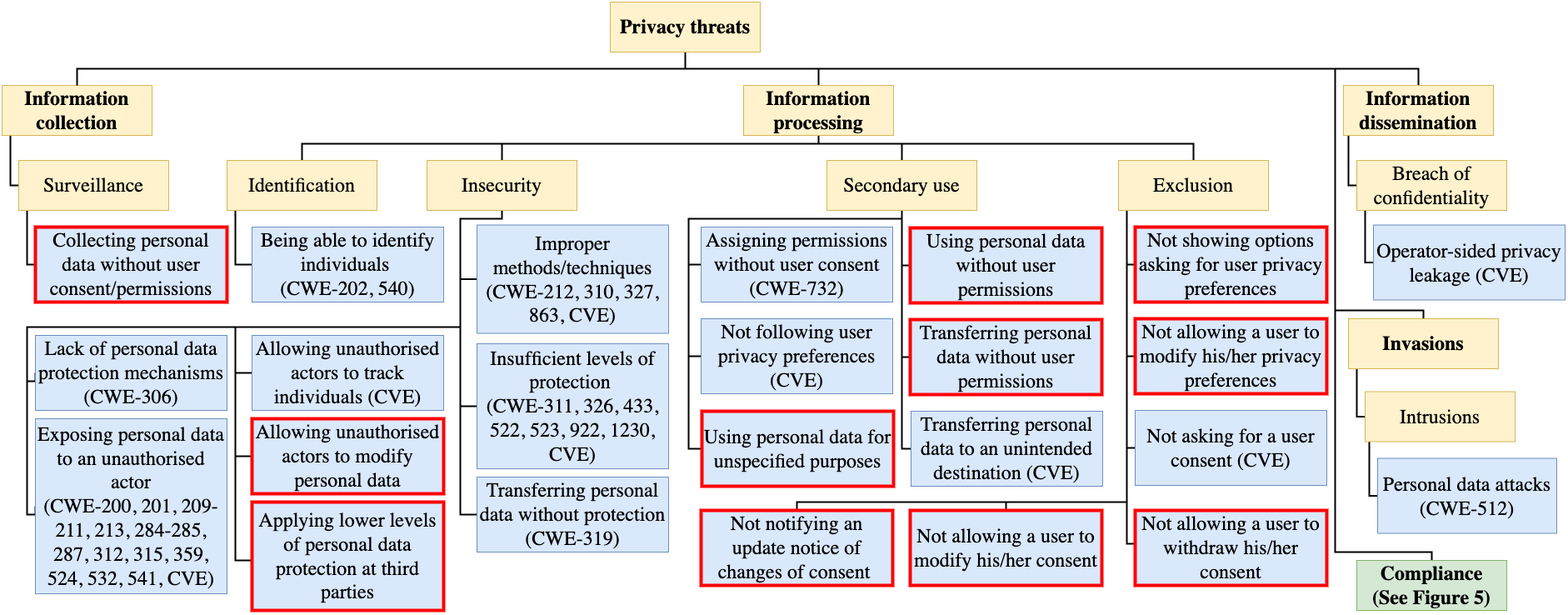}
	\caption{Common privacy vulnerabilities in software applications}
	\label{fig:AOC-vulnerabilities}
\end{figure*}

We have built a taxonomy of common privacy threats upon the well-established privacy threats taxonomy described in \cite{Stallings2019}. The taxonomy was originally proposed by \cite{Solove2006a}. It covers privacy violations of individuals that are caused by physical activities (e.g. a newspaper reports the name of a rape victim, or a company sells its members' personal information despite promising not to do so). Later, \citeauthor{Stallings2019} adapted the concept from \cite{Solove2006a} to build the taxonomy of privacy threats for information systems. However, Stallings's taxonomy covers privacy threats in a generic and rather abstract level. Thus, we tailored this taxonomy and refined it into a more concrete version that addresses privacy threats in SE. The taxonomy consists of four categories of privacy threats: information collection, information processing, information dissemination and invasions (see Figure \ref{fig:AOC-vulnerabilities}). Each of these groups contains different subcategories covering relevant harmful privacy threats. The yellow boxes in Figure \ref{fig:AOC-vulnerabilities} represent the categories and subcategories included in the original taxonomy.

\begin{figure}[ht]
	\centering
	\includegraphics[width=\linewidth]{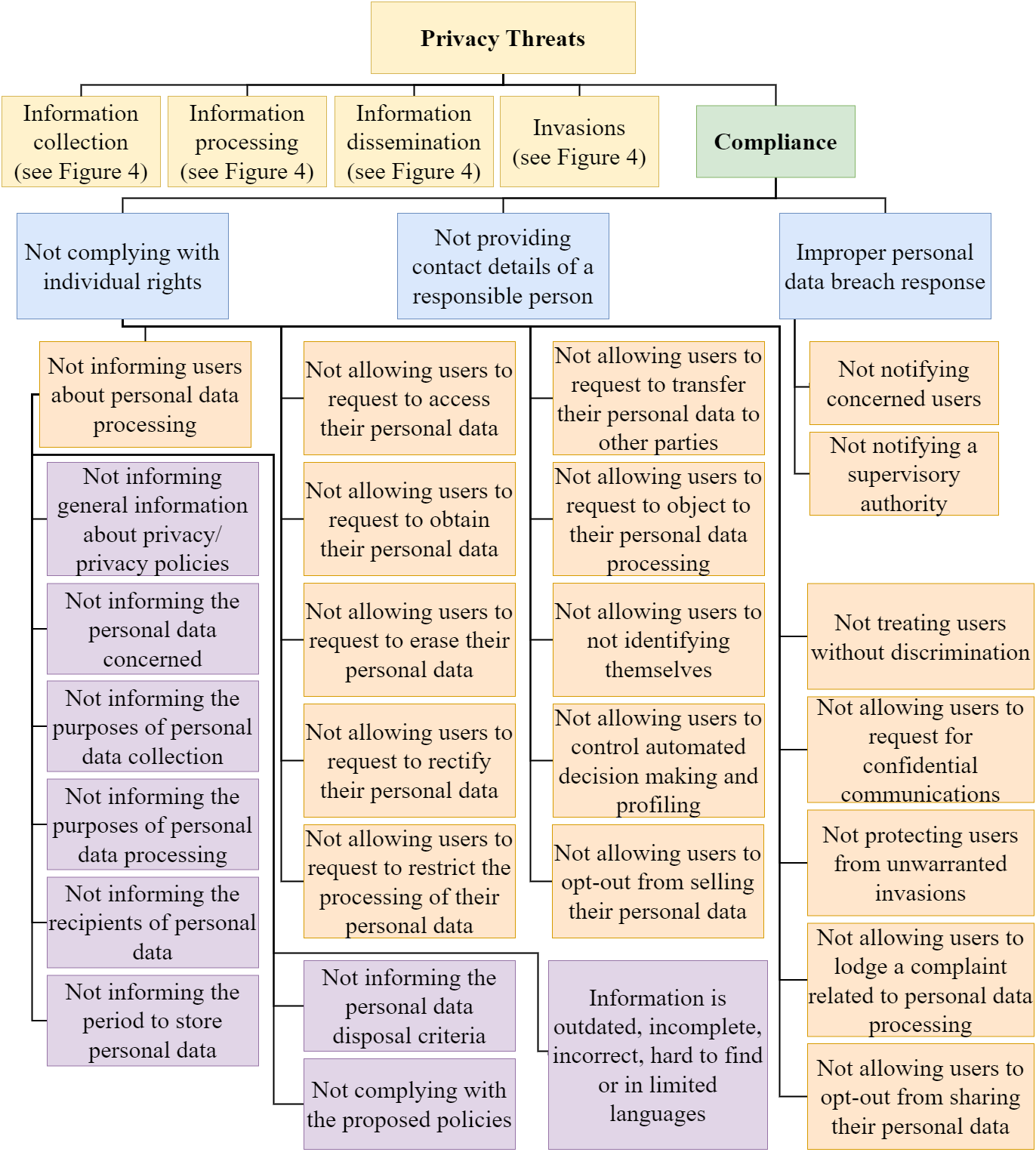}
	\caption{Common privacy compliance in software applications}
	\label{fig:AOC-compliance}
\end{figure}

After identifying the privacy threats in the explanatory study, we classified those threats into two groups: vulnerabilities and compliance. Privacy vulnerabilities refer to technical issues, flaws or errors that lead to privacy exploits in software applications. Compliance addresses the privacy threats related to the individual rights and the governance of personal data. We then expanded the Stallings's taxonomy by mapping the privacy threats in the vulnerabilities group into their relevant subcategories in the original taxonomy (i.e. blue boxes in Figure \ref{fig:AOC-vulnerabilities}). However, the privacy threats in the compliance group have not been addressed in any groups in the original taxonomy. Thus, we propose the compliance group as an extension to the original taxonomy (see Figure \ref{fig:AOC-compliance}). This group is discussed in details in Section \ref{subsec:compliance}. The full taxonomy is available at \cite{rep-pkg-privul}.

In our taxonomy, we classified 24 privacy vulnerabilities into 7 subcategories. In the classification process, we mapped a privacy threat into the most relevant subcategory based on the description of each subcategory explained in \cite{Stallings2019}. The categories, subcategories and their relevant privacy threats are described below. We also provide several examples of sources where the privacy threats were raised or discussed in each category\footnote{See the file named \emph{``4-RQ2-explanatory-study-papers''} in the data folder in the replication package for the full list of papers with their associated privacy threats.}.

\subsubsection{\textbf{Information collection}} (Sources: \cite{De2016, Lebeck2018, Hasan2020, OfficeJournaloftheEuropeanUnion;2016, ISO/IEC2011, OWASPsurvey}) This category concerns privacy threats that occur when collecting personal data from individuals. The \emph{surveillance} subgroup covers vulnerabilities existing in the way software collects personal data such as watching individuals through cameras or CCTVs, listening to individuals, or recording individuals' activities. The privacy threat related to this subgroup occurs when personal data is collected without consent or permissions from individuals (e.g. via mobile sensors).

\subsubsection{\textbf{Information processing}} (Sources: \cite{Deng2011, Fisk2015, Figueiredo2017, Venkatadri2018, Some2019, Zhang2020a, HIPAA, GLBA}) This category covers vulnerabilities related to the use, storage, and manipulation of the collected personal data. It contains four subgroups: identification, insecurity, secondary use and exclusion. The \emph{identification} subcategory addresses a privacy threat that aggregates personal data from various sources and uses it to identify individuals.

The vulnerabilities in \emph{insecurity} subcategory are caused by improper protection and handling of personal data. There are multiple forms of this vulnerability type such as lacking mechanisms to protect personal data, allowing unauthorised actors to access or modify personal data, track individual users and transferring personal data without protection. Personal data is sometimes required to be processed at third parties. This poses a privacy vulnerability since the third parties may apply weaker level personal data protection than the source does. In addition, this subcategory refers to the vulnerabilities where mechanisms to protect personal data are in place but they are not appropriate. Different types of personal data require different methods/techniques and levels of protection. Thus, appropriate protection mechanisms should be used to protect personal data against potential risks. This vulnerability type is often refined into improper techniques/methods and insufficient levels of protection (e.g. weak encryption).

The \emph{secondary use} subcategory refers to the use of personal data for other purposes without consent or not following user privacy preferences. Personal data used or transferred without user permissions, or to an unintended destination is a privacy vulnerability. The personal data used without following user privacy preferences can also cause a privacy vulnerability\footnote{For example, CVE-2005-2512 reports a vulnerability which could result in a privacy leak in mail.app in Mac OS 10.4.2 and earlier in which remote images are loaded against the user's preferences when an HTML message is printed or forwarded.}.

The \emph{exclusion} subcategory refers to the failure to provide individuals with notice and input for managing their personal data. These vulnerabilities relate to consent which allows the users to express their agreement on the use of their personal data in software applications. We note that consent handling may seem to be a part of compliance, however this subcategory focuses on the malfunctions that cause vulnerabilities in consent handling. User consent is required when the processing of personal data is not required by laws. Users should also be notified when the conditions on the consent are changed. In addition, users should be allowed to modify or withdraw their consent. One example of this privacy vulnerability is in mobile applications where users allow a specific permission to an app, however the permission is overridden in other apps without their consent \cite{Calciati, Zhang2020a}. Apart from user consent, user privacy preferences are also important. Privacy preferences enable users to personalise how they prefer their personal data to be managed. This privacy vulnerability type can be in two forms: privacy preferences not provided to the users, and users not be able to modify their privacy preferences.

\subsubsection{\textbf{Information dissemination}} (Sources: \cite{Calciati, Jana2013, Lucia2012, Zhang2020a, OWASPsurvey}) This category refers to the privacy threats that lead to the disclosure of personal data to public. The \emph{breach of confidentiality} subcategory covers the vulnerabilities that cause personal data leakage by those who are responsible for personal data processing (e.g. software developers, data processors and third parties).

\subsubsection{\textbf{Invasions}} (Sources: \cite{Deng2011, Reinheimer2020, OWASPsurvey, Nortona}) This category addresses attacks that directly affect individuals. The \emph{intrusions} subcategory covers vulnerabilities exploited by common privacy attacks in software applications. There are typically four attacks: web applications, phishing, keystroke logging and smart home devices. 

\subsubsection{\textbf{Compliance}} \label{subsec:compliance}

We used a bottom-up approach to construct a taxonomy for the compliance category. The privacy threats that address the same concerns are grouped into the same subcategory. We classify privacy compliance into the following three subcategories. The first is \emph{not complying with individual rights} (Sources: \cite{OfficeJournaloftheEuropeanUnion;2016, CCPA, HIPAA, GLBA, US1974, APA, ISO/IEC2011, Deng2011, Omoronyia2013a, Bhatia2018a, Yu2021, Mihaylov2016a, OWASPsurvey}). There are 15 individual rights identified in our study\footnote{See the file named \emph{``14-rights''} in the data folder in the replication package for the complete list of individual rights.}. 

The second is \emph{not providing contact details of a responsible person} (Sources: \cite{OfficeJournaloftheEuropeanUnion;2016, ISO/IEC2011}). This subcategory covers cases where software applications do not provide the contact details of a responsible person or a representative who processes personal data. This is a privacy threat as the users do not know whom to be contacted regarding their personal data processing. The third is \emph{improper personal data breach response} (Sources: \cite{OfficeJournaloftheEuropeanUnion;2016, ISO/IEC2011, HIPAA, OWASPsurvey}). When a breach occurs, a responsible person must notify two key stakeholders: concerned individuals and a supervisory authority. The privacy threat is raised if the responsible person does not communicate the breach incident to the concerned users whose personal data is leaked to the public and the supervisory authority who monitors the personal data processing under the individual rights.

Our taxonomy covers the existing vulnerabilities occurred in a range of activities performed in software systems (e.g. personal data collection and processing). These privacy vulnerabilities have been raised in real world software applications and software development processes. The taxonomy was also developed based on an existing comprehensive privacy threats taxonomy, and validated with the common vulnerabilities reported in CWE and CVE.

\subsection{Privacy threats covered in CWE/CVE} \label{subsec:addressing-privacy-concerns}

The next step in our study was to investigate how 41 privacy vulnerabilities in CWE and 157 in CVE (see Section \ref{sec:identifying-privacy-vul}) address the taxonomy of common privacy threats in Section \ref{subsec:areas-of-concern}. This process consists of the following steps. The first two co-authors (hereafter referred to as the coders) \emph{independently} analysed each of the 41 privacy vulnerabilities in CWE, and classified it into the most relevant privacy threat in the taxonomy. Of the 157 privacy vulnerabilities in CVE, 112 were assigned to a specific CWE by the NVD. These 112 CVEs are automatically classified into the same privacy threat as their associated CWE. The remaining 45 CVEs, which were not assigned a specific CWE\footnote{NVD used two special placeholder names for these: NVD-CWE-noinfo and NVD-CWE-other.}, were manually classified by our coders. To facilitate the classification step, each coder was provided with a Google Sheet form pre-filled with the privacy vulnerabilities in CWE and CVE and the privacy threats in the taxonomy. The privacy threats were prepared as a drop down list. The coders then selected the vulnerability that is most relevant and best described those CWE and CVE in their view.

We have also employed the Inter-Rater Reliability (IRR) assessment and disagreement resolution processes to ensure the reliability of the classifications. The Cohen's Kappa coefficient is used to measure the inter-rater agreement as it is a perfect measure for a multi-class classification problem with two coders \cite{Hallgren}. Once the coders had completed the classifications, the inter-rater agreement was computed. The Kappa agreement values between the coders are 0.874 and 0.875 in CWE and CVE classifications respectively, which both achieve \emph{almost perfect agreement} level \cite{Viera2005}. A disagreement resolution was conducted to resolve some small classification conflicts (4 CWEs and 4 CVEs). The coders met, went through them together, discussed and reclassified those vulnerabilities. Thus, the final classification reached the maximum agreement between the coders.

\subsubsection{Results}

We have found that all the 41 CWEs and 157 CVEs together cover 13 vulnerabilities in the taxonomy. They are annotated with the corresponding CWEs and CVEs in Figure \ref{fig:AOC-vulnerabilities}. For brevity, we do not include the CVE numbers in Figure \ref{fig:AOC-vulnerabilities}, but they are provided in full in our replication package \cite{rep-pkg-privul}. Exposing personal data to an unauthorised actor, insufficient levels of protection, and personal data attacks are the top three most addressed vulnerabilities in both CWE and CVE. Personal data protection seems to attract a lot of attentions in CWE/CVE with more than half of the CWEs (56.1\%) and CVEs (59.87\%) vulnerabilities reported, most of which (36.59\% in CWEs and 50.32\% in CVEs) are related to exposing personal data to an unauthorised actor. There are 19.51\% in CWE and 15.92\% in CVE reporting vulnerabilities regarding personal data attacks.

There are four types of privacy vulnerabilities that are covered by both CWE and CVE: exposing personal data to an unauthorised actor, insufficient levels of personal data protection, improper methods/techniques for personal data protection, and personal data attacks. There are several types of privacy vulnerabilities that have been reported in CVE, but not in CWE, such as allowing unauthorised actors to track individuals, not following user privacy preferences, and not asking for user consent. 8.28\% of the CVEs refer to those types of privacy vulnerabilities, suggesting that those types of vulnerabilities need to be added into the CWE system.

A number of areas that are not covered by the existing privacy vulnerabilities in CWE and CVE are highlighted in red outline in Figure \ref{fig:AOC-vulnerabilities}. For example, exclusion involves a range of privacy vulnerabilities in failing to provide users with notice of user consent and input about their privacy preferences. User consent and privacy preferences are two essential mechanisms that enable users to control their personal data processing. These sources \cite{Antn2004, ISO/IEC2011, HIPAA} confirm that user privacy is vulnerable if users are not presented with options to specify or cannot modify their user privacy preferences. Similarly, user privacy may be  compromised if the users cannot modify or withdraw their consent, or are not notified about any changes of consent \cite{OfficeJournaloftheEuropeanUnion;2016, ISO/IEC2011, HIPAA, APA, OWASPsurvey}. However, none of them (except not asking for a user consent vulnerability) is covered in existing CVE.

Other areas that are not well covered in existing CWEs and CVEs are privacy vulnerabilities due to insecurity and secondary use. Processing personal data at a third party is a risk since they may apply weaker level personal data protection, particularly mobile \cite{Zhang2020a} and web applications \cite{OWASPsurvey}. There are cases that user privacy is violated when personal data is used for unspecified purposes or used or transferred without permissions (e.g. \cite{Hasan2020, De2016, Fisk2015, HIPAA, GLBA, US1974}). In addition, although allowing unauthorised actors to modify personal data and collecting personal data without user consent/permissions are serious threats, none of the existing privacy vulnerabilities in CWE and CVE covers this.

\begin{conclusion}
	\textbf{The existing privacy weaknesses and vulnerabilities reported in the CWE and CVE systems cover only 13 out of 24 common privacy vulnerabilities raised in research and practice.}
\end{conclusion} 
\section{New common privacy weaknesses} \label{sec:cwe-proposal}

Our study has shown the gaps in the existing CWE and CVE systems in terms of covering privacy vulnerabilities. To fill these gaps, we propose 11 new common privacy weaknesses to be added to the CWE system. We focus on CWE instead of CVE since CVE specifies unique vulnerabilities existing in specific software systems and application, while CWEs are at the generic level, similar to our taxonomy of privacy threats. We followed the CWE schema \cite{CWEschema2021} to define the new CWE entries which include attributes such as name, description, mode of introduction, common consequence, detection method, potential mitigation and demonstrative example. These attributes provide an overview of a privacy weakness in terms of its causes and consequences, and mitigation methods.

\newtext{Security vulnerabilities can typically be well specified as concrete hardware/software attacks that can be exploited to bypass the system's security. Unlike security vulnerabilities, privacy vulnerabilities seem to be more contextual than security vulnerabilities and its operationalisation may differ depending on context. In our study, we tried to formalise the privacy vulnerabilities as software attacks that are harmful to personal data and violate individual rights. Thus, it is important to provide the detailed description in each attribute in the identified 11 common privacy weaknesses since it defines a specific scenario that potentially leads privacy vulnerabilities.}

The proposed common privacy weaknesses address four groups of threats: surveillance, insecurity, secondary use and exclusion. The weaknesses in the insecurity group can be detected and resolved by implementing security mechanisms to better protect personal data and user privacy. On the other hand, the other groups of weaknesses require more attention from software development teams on examining privacy constraints involved, and designing relevant functions to respond to those constraints. The software development team needs to determine relevant functions, take privacy constraints into consideration and regularly review existing functions to ensure that they do not violate user privacy. For example, a user may want to change his/her user preferences when a new feature is launched, or a system must notify its users when there is a change to the policy that they have given consent to.

Due to space limit, we present here two examples (see Table \ref{tab:cwe-template-missing-consent-withdrawal} and Table \ref{tab:cwe-template-missing-consent-check}) of the new CWE privacy weaknesses and refer to the readers to our replication package \cite{rep-pkg-privul} for the remaining newly proposed CWEs. Table \ref{tab:cwe-template-missing-consent-withdrawal} shows a new common privacy weakness for not allowing a user to withdraw his/her consent. This weakness belongs to the exclusion category. Following the CWE schema, a short summary of the weakness is provided in its name, while a detailed description is provided in the description section. The mode of introduction briefly discusses how and when the weakness is introduced, which in this case is the architecture and design phase.

\begin{table}[ht]
	\centering
	\caption{An example of a missing consent withdrawal weakness.}
	\label{tab:cwe-template-missing-consent-withdrawal}
	\begin{tabular}{|p{8.5cm}|}
		\hline
		\textbf{Subcategory:} Exclusion \\
		\textbf{Class:} Not allowing a user to withdraw his/her consent \\
		\textbf{Name:} Missing consent withdrawal  \\
		\textbf{Description:} The software forces users to give consent before providing its services. These services may include personal data processing. However, the software does not a provide a function for users to withdraw their consent. The users can only accept consent, but they cannot withdraw their consent when they wish to. This vulnerability seriously violates user privacy. \\
		\textbf{Mode of introduction:} Phase: Architecture and Design. This weakness is caused by a missing privacy consideration about consent management and its related processes, which leads to the missing consent withdrawal function. \\
		\textbf{Common consequence:} The software violates user privacy by not allowing users to express their agreement on the use of their personal data. \\
		\textbf{Detection method:} Method: Manual analysis. Description: A consent management page or window in a software does not show an icon or option to withdraw consent. \\
		\textbf{Potential mitigations:} Phase: Architecture and Design. Strategy: User consent withdrawal consideration. Description: The software development team should consider which points in the software that should provide a user an ability to withdraw consent. \\
		\textbf{Demonstrative example:} Issue MDL-62309 in Moodle reports that the users cannot withdraw consent and cannot enter the site without giving consent. This issue violates user privacy as the users should be able to freely withdraw consent. \\
		\hline
	\end{tabular}
\end{table}

\begin{figure}[ht]
	\centering
	\includegraphics[width=1.0\linewidth]{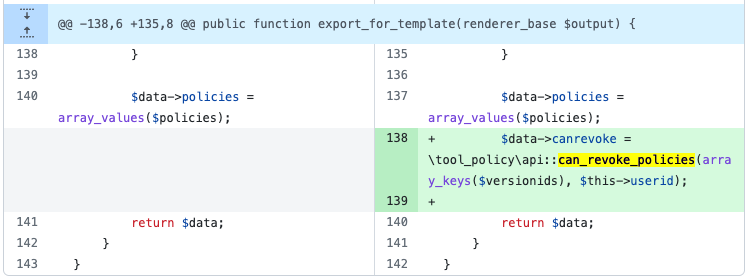}
	\caption{An example of missing consent withdrawal function in Moodle (Issue MDL-62309).}
	\label{fig:mdl-missing-consent-withdrawal}
\end{figure}

\begin{table}[ht]
	\centering
	\caption{An example of a collecting personal data without user consent/permissions weakness.}
	\label{tab:cwe-template-missing-consent-check}
	\begin{tabular}{|p{8.5cm}|}
		\hline
		\textbf{Subcategory:} Surveillance \\
		\textbf{Class:} Collecting personal data without user consent/permissions \\
		\textbf{Name:} Missing a consent check before collecting personal data \\
		\textbf{Description:} The software does not check for a user consent prior to personal data collection. This makes the software collects personal data that users have not given consent to (e.g., location and speech). \\
		\textbf{Mode of introduction:} Phase: Implementation. This weakness is caused by missing a consent check before collecting personal data. \\
		\textbf{Common consequence:} The software violates user privacy since users has not given consent/permissions to collect their personal data. \\
		\textbf{Detection method:} Method: Manual analysis. Description: Perform a code check at points of personal data collection. \\
		\textbf{Potential mitigations:} Phase: Implementation. Strategy: Check for user consent before collecting data. Description: The software development team should perform a consent check at every point that collects personal data in the software. \\
		\textbf{Demonstrative example:} A commit 0b09df0 in HumanDynamics repository collects user speech without consent check. \\
		\hline
	\end{tabular}
\end{table}

\begin{figure}[ht]
	\centering
	\includegraphics[width=1.0\linewidth]{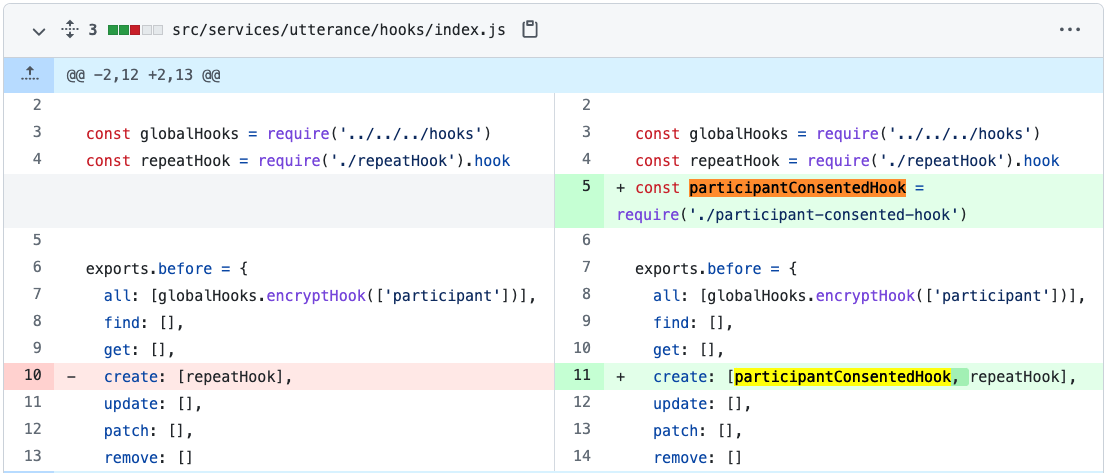}
	\caption{An example of missing a consent check before collecting personal data in HumanDynamics (Commit 0b09df0).}
	\label{fig:hd-missing-consent-check}
\end{figure}

The common consequence section identifies a privacy property that is violated and an effect that is caused by the weakness. The detection method section describes different methods that the weakness can be detected in software. We also propose methods to mitigate the weakness. It is noted that different phases in software development may pose different privacy concerns. Finally, we provide a demonstrative example of the new weaknesses by extracting code fragments, issue reports and commits from real software repositories hosted on GitHub\footnote{\url{https://github.com/}}. \newtext{We used the keywords in the weakness class and name to search for relevant code, commits and issues. Once the search results were returned from GitHub, we went through the descriptions and identified the demonstrative example from those results.} For example, Figure \ref{fig:mdl-missing-consent-withdrawal} shows a code fragment extracted from the Github commit \emph{ad5e213}\footnote{\url{https://github.com/moodle/moodle/commit/ad5e213}} in Issue MDL-62309\footnote{https://tracker.moodle.org/browse/MDL-62309} of the Moodle project. This example demonstrates the existence of the missing consent withdrawal weakness in practice.

Table \ref{tab:cwe-template-missing-consent-check} presents a new common privacy weakness for collecting personal data without user consent/permissions. This weakness belongs to improper personal data collection category. Figure \ref{fig:hd-missing-consent-check} shows a code fragment confirming the existence of a missing consent check before collecting personal data weakness. The code fragment is extracted from the Github commit \emph{0b09df0}\footnote{\url{https://github.com/HumanDynamics/rhythm-server/commit/0b09df0}} in the HumanDynamic repository \cite{HumanDynamics}.

\begin{conclusion}
	\textbf{We propose 11 new common privacy weaknesses to be added to CWE. These will significantly improve the coverage of privacy weaknesses and vulnerabilities in CWE, and subsequently CVE.}
\end{conclusion}

\section{Threats to validity} \label{sec:threats}

\textit{\textbf{Internal validity.}} Our method for the extraction of privacy vulnerabilities in CWE and CVE using keywords might not result in the complete list. We also note that we did not consider the variations of the selected keywords (e.g. plural forms). However, we have used several strategies to mitigate these threats such as determining the keywords based on alternate terms described in CWE, and using the frequent terms identified in the studies that performed a large-scale analysis in privacy policies and considering general terms to cover unseen materials (e.g. regulation, data protection and privacy standard). 

\textit{\textbf{External validity.}} Our taxonomy of common privacy threats is constructed and refined based on the existing privacy threats taxonomy. We have put our best effort to ensure the comprehensiveness of the study by examining popular software engineering publication venues, well-established data protection regulations and privacy frameworks, and relevant reputable organisations. However, we acknowledge that there might be other sources of other privacy threats that we have not identified yet. We have carefully defined a set of inclusion and exclusion criteria to select the most relevant papers so that we got a reasonable number of papers to be examined individually. Future work would involve expanding our explanatory study to increase the generalisability of our taxonomy for common privacy threats. In addition, classifying the privacy-related vulnerabilities in CWE and CVE into the common privacy threats in the taxonomy involved subjective judgements. We have applied several strategies (e.g. using multiple coders, applying inter-rater reliability assessments and conducting disagreement resolution) to mitigate this threat. Future work could explore the use of external subject matter experts in these tasks.

\section{Conclusions and future work} \label{sec:conclusion}

The increasing use of software applications in people's daily lives has put privacy under constant threat as personal data are collected, processed and transferred by many software applications. In this paper, we performed a number of studies on privacy vulnerabilities  in software applications. Our study on CWE and CVE systems found that the coverage of privacy-related vulnerabilities in both systems is quite limited (4.45\% in the CWE system and 0.1\% in the CVE system).

We have also investigated on how those privacy-related vulnerabilities identified in CWE and CVE address the common privacy threats in software applications. To do so, we developed a taxonomy of common privacy threats, which was extended from the existing work \cite{Stallings2019}, based on selected privacy engineering research, data protection regulations and privacy frameworks and industry resources. 
We have found that only 13 out of 24 common privacy vulnerabilities in the taxonomy are covered by the existing weaknesses and vulnerabilities reported in CWE and CVE. 
Based on these actionable insights, we proposed 11 new common privacy weaknesses to be added to the CWE system. We also mined code fragments from real software repositories to confirm the existence of those privacy weaknesses. These newly proposed weaknesses will significantly improve the coverage of privacy weaknesses and vulnerabilities in CWE, and subsequently CVE.

Future work involves expanding our taxonomy to cover additional common privacy threats that may have raised or discussed in other sources. We will also perform a study to characterise privacy vulnerabilities in software applications. This will enable us to develop new techniques and tools for automatically detecting privacy vulnerabilities in software and suggesting fixes.

\section{Data Availability} \label{sec:data-availability}

A full replication package containing all the artifacts and data produced by our study is made publicly available at \url{https://figshare.com/articles/conference_contribution/icse2023-paper908-replication-pkg/21922731} or DOI: \url{https://doi.org/10.6084/m9.figshare.21922731}. There are five folders in the package: i) accepted-paper, ii) code, iii) common-weakness-proposal, iv) data and v) taxonomy. The \textbf{README} file that explains relevant folders and files is also included in the replication package.

\balance
\bibliographystyle{IEEEtranN}
\bibliography{Privacy-vulnerabilities}

\end{document}